\input harvmac
\Title{\vbox{\baselineskip12pt
\hbox{CERN-TH/96-235}
\hbox{McGill/96-32}
\hbox{hep-th/9609094}
}}
{\vbox{\centerline{Supersymmetry, Duality and Bound States}}}
\centerline{Ramzi R.~Khuri\foot{Talk given at ``Strings '96'',
Institute
for Theoretical Physics, University of California at Santa Barbara,
July 15-20, 1996. Supported by a World Laboratory Fellowship.}}
\bigskip\centerline{{\it Theory Division, CERN,
 CH-1211, Geneva 23, Switzerland}}
\bigskip\centerline{and}
\bigskip\centerline{{\it Physics Department, McGill
University, Montreal, PQ, H3A 2T8 Canada}}
\vskip .3in

$P$-brane solutions of low-energy string actions have traditionally
provided the first evidence for the existence of string dualities, in
which fundamental and solitonic $p$-branes are identified with
perturbative and non-perturbative BPS states. In this talk we discuss
the composite nature of solutions, which allows for the
interpretation of general solutions as bound states or intersections
of maximally supersymmetric fundamental constituents. This feature
lies
at the heart of the recent success of string theory in reproducing
the
Beckenstein-Hawking black hole entropy formula.

\vskip .3in
\Date{\vbox{\baselineskip12pt
\hbox{CERN-TH/96-235}
\hbox{McGill/96-32}
\hbox{September 1996}}}

\def\sqr#1#2{{\vbox{\hrule height.#2pt\hbox{\vrule width
.#2pt height#1pt \kern#1pt\vrule width.#2pt}\hrule height.#2pt}}}
\def\Box{\mathchoice\sqr64\sqr64\sqr{4.2}3\sqr33}

\def\[{\bf (*}
\def\]{*) \rm\ }

\lref\kal{D.~J.~Gross and M.~J.~Perry, Nucl. Phys. {\bf B226}
(1983) 29; R.~D.~Sorkin, Phys. Rev. Lett. {\bf 51} (1983) 87.}

\lref\duflufb{M.~J.~Duff and J.~X.~Lu, Nucl. Phys. {\bf B354} (1991)
141.}

\lref\antft{I.~Antoniadis, S.~Ferrara and T.~R.~Taylor, Nucl. Phys.
{\bf B460} (1996) 489.}

\lref\duffst{M.~J.~Duff and K.~S.~Stelle, Phys. Lett. {\bf B253}
(1991) 113.}

\lref\guven{R.~G\" uven, Phys. Lett. {\bf B276} (1992) 49.}

\lref\iib{J.~H.~Schwarz, Phys. Lett. {\bf B360} (1995) 13-18,
ERRATUM-ibid. {\bf B364} (1995) 252.}

\lref\fat{J.~M.~Maldacena and L.~Susskind, hep-th/9604042;
E.~Halyo, A.~Rajaraman and L.~Susskind, hep-th/9605112.}

\lref\hmono{R.~R.~Khuri, Phys. Lett. {\bf B259} (1991) 261;
Nucl. Phys. {\bf B387} (1992) 315.}

\lref\jkkm{C.~V.~Johnson, N.~Kaloper, R.~R.~Khuri and R.~C.~Myers,
Phys. Lett. {\bf B368} (1996) 71.}

\lref\joe{J.~Polchinski, Phys. Rev. Lett. {\bf 75} (1995)
4724.}

\lref\pcj{J.~Polchinski, S.~Chaudhuri and C.~V.~Johnson,
hep-th/9602052.}

\lref\hult{C.~M.~Hull and P.~K.~Townsend, Nucl. Phys. {\bf B438}
(1995) 109.}

\lref\ed{E.~Witten, Nucl. Phys. {\bf B443} (1995) 85.}

\lref\prep{M.~J.~Duff, R.~R.~Khuri and J.~X.~Lu,
Phys. Rep. {\bf 259} (1995) 213.}

\lref\stst{M.~J.~Duff and R.~R.~Khuri, Nucl. Phys. {\bf B411} (1994)
473; M.~J.~Duff, Nucl. Phys. {\bf B442} (1995) 47.}

\lref\dufflm{M.~J.~Duff, J.~T.~Liu and R.~Minasian, Nucl. Phys.
{\bf B452} (1995) 261.}

\lref\schwarz{J.~H.~Schwarz, Phys. Lett. {\bf B367} (1996) 97.}

\lref\horw{P.~Ho\v rava and E.~Witten, Nucl. Phys. {\bf B460} (1996)
506.}

\lref\rust{R.~R.~Khuri and R.~C.~Myers, Nucl. Phys. {\bf B466} (1996)
60; hep-th 9512137.}

\lref\bifb{R.~R.~Khuri, Phys. Rev. {\bf D48} (1993) 2947.}

\lref\dfkr{M.~J.~Duff, S.~Ferrara, R.~R.~Khuri and J.~Rahmfeld,
 Phys. Lett. {\bf B356} (1995) 479.}

\lref\fkm{S.~Ferrara, R.~R.~Khuri and R.~Minasian,
 Phys. Lett. {\bf B375} (1996) 81.}

\lref\bmike{M.~J.~Duff and J.~Rahmfeld, Phys. Lett. {\bf B345} (1995)
441.}

\lref\bjoachim{J.~Rahmfeld Phys. Lett. {\bf B372} (1996) 198;
M.~J.~Duff and J.~Rahmfeld, hep-th/9605085.}

\lref\bed{E.~Witten, Nucl. Phys. {\bf B460} (1996) 335.}

\lref\bsusy{R.~R.~Khuri and T.~Ortin, Nucl. Phys. {\bf B467} (1996)
355.}

\lref\bchris{ N.~Khviengia, Z.~Khviengia, H.~Lu and C.~N.~Pope,
hep-th/9605077; H.~Lu and C.~N.~Pope, hep-th/9606047.}

\lref\bmirjam{M.~Cveti\v c and D.~Youm, Phys. Rev. Lett. {\bf 75}
(1995) 4165; Nucl. Phys. {\bf B453} (1995) 259; Phys. Lett. {\bf
B359} (1995) 87; Phys. Rev. {\bf D53} (1996) 584.}

\lref\barkady{M.~Cveti\v c and A.~A.~Tseytlin, Phys. Rev. {\bf D53}
(1996) 5619; Phys. Lett. {\bf B366} (1996) 95; A.~A.~Tseytlin,
gr-qc/9608044.}

\lref\iarkady{A.~A.~Tseytlin, hep-th/9604035;
I.~R.~Klebanov and A.~A.~Tseytlin, hep-th/9604166.}

\lref\iklaus{K.~Behrndt, E.~Bergshoeff, B.Janssen, hep-th/9604168;
K.~Behrndt and E.~Bergshoeff, hep-th/9605216.}

\lref\ijerome{J.~P.~Gauntlett, D.~A.~Kastor and J.~Traschen,
hep-th/9604179.}

\lref\imirjam{M.~Cveti\v c and A.~A.~Tseytlin, hep-th/9606033.}

\lref\exfive{C.~Vafa and A.~Strominger, Phys. Lett. {\bf B379}
(1996) 99; S.~R.~Das and S.~D.~Mathur, Phys. Lett. {\bf B375}
(1996) 103; C.~G.~Callan and J.~M.~Maldacena, Nucl. Phys. {\bf B472}
(1996) 591.}

\lref\exfour{J.~M.~Maldacena and A.~Strominger, Phys. Rev. Lett.
{\bf 77} (1996) 428; C.~V.~Johnson, R.~R.~Khuri and R.~C.~Myers,
Phys. Lett. {\bf B378} (1996) 78.}

\lref\nonex{G.~T.~Horowitz, J.~M.~Maldacena and A.~Strominger,
hep-th/9603109; G.~T.~Horowitz, D.~A.~Lowe and J.~M.~Maldacena,
Phys. Rev. Lett. {\bf 77} (1996) 430; J.~M.~Maldacena,
hep-th/9605016.}


In the last few years, it has become increasingly clear that
$p$-branes, i.e. $(p{+}1)$-dimensional extended objects that
arise as fundamental or solitonic solutions of string theory,
play an important role in establishing the various string dualities
and ultimately in the non-perturbative formulation of an underlying
theory containing strings and higher membranes.

Under a given duality map, the fundamental $p$-brane
of a given string or higher membrane theory tranforms into
a solitonic solution of the dual theory, with the corresponding
interchange of singular/non-singular backgrounds,
perturbative/non-perturbative states and even spacetime/worldsheet
loops (see \prep\ and references therein).

An interesting
example of the implications of duality is that of six-dimensional
heterotic string/string duality, which, when reduced to
four dimensions, leads to a {\it duality of
dualities}, namely, the interchange of the spacetime strong/weak
coupling $S$-duality with the worldsheet target space $T$-duality
\stst.
As a result, the conjectured $S$-duality follows from the
established $T$-duality
provided four-dimensional string/string duality is shown.

More recently, interest has focused on duality between different
string theories, in particular, heterotic/type IIA string/string
duality in $D=6$ \refs{\hult,\ed}. In this case, one has a more
straightforward strong/coupling duality relating the two
theories with $g_{IIA}=1/g_{het}$.

The presence of certain $p$-branes, however, coupled with the
assumption of
a given duality, can point to gaps in the formulation of string
theory.
In particular, it was shown in \jkkm\ that, in the context of
heterotic/type duality,
aside from the usual interchange of fundamental and solitonic
strings,
there exist membrane solutions which are non-singular (solitonic)
in the heterotic theory but singular in the type IIA theory. Such
solutions were then interpreted as fundamental membranes in type IIA,
since
they could neither be ignored as purely singular configurations nor
simply counted in the solitonic spectrum of either theory. The
authors
of \jkkm\ then concluded that the formulation of type IIA theory as
a theory of strings alone was incomplete. The subsequent discovery of
the role of $D$-branes as carriers of Ramond-Ramond (RR)
charge that should be
coupled to type II theories \joe\
(see also \pcj) effectively resolved this
problem, since the fundamental membrane solution of \jkkm\
could now have the interpretation of a $D$-brane.

A picture then emerges in which fundamental and solitonic
$p$-branes carrying Neveu-Schwarz-Neveu-Schwarz charge
correspond to perturbative and non-perturbative BPS states of
string theory, while $p$-branes carrying RR charge correspond
to $D$-brane BPS states which must be coupled to the perturbative
spectrum. From both the $p$-brane and BPS state points of view,
the mass of a fundamental solution (state) is
independent of the string coupling $g$, while
the mass of a solitonic solution (state) scales as
$1/g^2$. By contrast, the mass of a $D$-brane scales as $1/g$
\refs{\iib,\bed}.
This intermediate behaviour
between fundamental and solitonic states is an inherently stringy
departure from the usual $1/g^2$ solitonic scaling found in field
theory.

Recent activity has also focused on the existence of an
eleven-dimensional
theory, the so-called $M$-theory \refs{\ed,\dufflm\schwarz{--}\horw},
whose low-energy limit is eleven-dimensional supergravity and whose
construction should lead to the establishment of the various
string/string
dualities \refs{\stst,\hult,\ed}. While we still do not know exactly
what $M$-theory is, we know that
$M$-theory contains membranes and fivebranes. From the point of
view of $p$-brane solutions, these are represented by a fundamental
membrane \duffst\ and a solitonic fivebrane \guven.

An example of how $p$-brane solutions can provide evidence for
a duality is the following. In \antft, the conjecture was made
that the effective theory of heterotic string theory compactified
on $K3\times S^1$ is dual to
eleven-dimensional supergravity compactified on a Calabi-Yau
threefold.
Point-like (electric) solutions are obtained in $D=5$ by wrapping the
membrane from $M$-theory around two-cycles in
the Calabi-Yau space, while
string-like (magnetic) solutions in $D=5$ arise by wrapping the
fivebrane around four-cycles.
For the specific Calabi-Yau manifold
$X_{24}(1,1,2,8,12)$ with $h_{(1,1)}=3$ and $h_{(2,1)}=243$, these
point and string solutions/states can be matched with
perturbative and non-perturbative solutions/states of
heterotic string theory compactified on $K3\times S^1$ \fkm.

The heterotic perturbative solutions (or states) include the
fundamental string and electrically charged
point-like $H$-monopole \hmono\
and Kaluza-Klein \kal\ solutions, the latter
two being obtained by wrapping the string around $S_1$.
The fundamental
string can be identified with one of the three states arising
from
the $M$-theory fivebrane, while the  electric $H$-monopole
and Kaluza-Klein
monopole
can be identified with two of the three states arising from the
$M$-theory membrane.

The non-perturbative heterotic solutions/states
consist of the point-like solution
obtained from the heterotic fivebrane wrapped around $K3\times S^1$
and the magnetically charged
string-like $H$-monopole and Kaluza-Klein solutions obtained by
wrapping the fivebrane around $K3$ only.
Here the point-like state can be identified with one of the three
states
arising from the $M$-theory membrane, while the
magnetic $H$-monopole and Kaluza-Klein states
can be identified with two of the three states
arising from the $M$-theory fivebrane.

The ``basic'' fundamental and solitonic $p$-branes preserve
half the spacetime supersymmetries, and arise
as extremal limits of more general, non-supersymmetric black
$p$-brane
solutions of string theory. It turns out, however, that the
low-energy
supergravity equations of motion possess a
feature that allows for the immediate construction of composite
solutions from the basic ones. Consider, for example, the
double-instanton string solution of heterotic string theory \bifb
\eqn\bifbsol{\eqalign{\phi&=\phi_1 + \phi_2,\cr
g_{mn}&=e^{2\phi_1}\delta_{mn}\qquad m,n=2,3,4,5,\cr
g_{ij}&=e^{2\phi_2}\delta_{ij}\qquad i,j=6,7,8,9,\cr
g_{\mu\nu}&=\eta_{\mu\nu}\qquad\quad   \mu,\nu=0,1,\cr
H_{mnp}&=\pm 2\epsilon_{mnpq}\partial^q\phi
\qquad m,n,p,q=2,3,4,5,\cr
H_{ijk}&=\pm 2\epsilon_{ijkl}\partial^k\phi
\qquad i,j,k,l=6,7,8,9\cr}}
with $e^{-2\phi_1}\Box_1 e^{2\phi_1}=e^{-2\phi_2}\Box_1
e^{2\phi_2}=0$,
where $\Box_1$ and $\Box_2$ are the Laplacians in the $(2345)$ and
$(6789)$
spaces, respectively.
For constant chiral spinors
$\epsilon=\epsilon_2 \otimes \eta_4 \otimes \eta_4'$, \bifbsol\
solves
the supersymmetry equations with zero background fermi fields. Due
to the presence of two independent instantons \hmono\ in
the generalized curvature containing $H_{MNP}$ as a torsion,
the chiralities of the spinors $\epsilon_2$, $\eta_4$ and $\eta_4'$
are correlated by
$(1 \mp \gamma_3)\epsilon_2=(1 \mp \gamma_5)\eta_4
=(1 \mp \gamma_5)\eta_4'=0$,
so that $1/4$ of the spacetime supersymmetries is preserved.

For either $\phi_1=0$ or $\phi_2=0$ we recover the solitonic
fivebrane
solution
of \duflufb\ which preserves $1/2$ the spacetime supersymmetries,
since
only one instanton is present. In this respect, this string
is the composite of two independent fivebranes intersecting
along the string. This feature can in fact be generalized to
arbitrary
$p$-brane solutions \refs{\iarkady\iklaus\ijerome{--}\imirjam},
whereby
a given $p$-brane soliton can be interpreted as the intersection of
one or more maximally supersymmetric basic fundamental or solitonic
$p$-branes. From the viewpoint of $M$-theory, the statement then
translates into saying that all $p$-brane solutions arise as
intersections of membrane and fivebranes \iarkady.

Compositeness was also seen in the specific
context of string-like solutions of toroidally compactified $D=4$
heterotic string theory \dfkr, where each solution could be
understood
in terms of four independent harmonic functions. The existence of
these
latter solutions also pointed to an interesting interplay between
supersymmetry and duality. In particular, the different supersymmetry
breaking patterns of the string-like solutions conform to the
different large duality groups (containing both $S$ and $T$ duality)
of the various compactifications \dfkr.

A related composite
picture of $p$-brane solutions is the bound states picture
\refs{\bmike\bjoachim\bed\bsusy\bmirjam\bchris{--}\barkady}.
For the simplest
case of extremal black hole solutions in $D=4$, consider the
Einstein-Maxwell scalar action
\eqn\amod{S=\int d^4x \left( R -{1\over 2}(\partial\phi)^2 - {1\over
4}
e^{-a\phi} F_2^2 \right),}
where $a$ is an arbitrary parameter. It turns out that for the
specific values of $a=\sqrt{3},1,1/\sqrt{3}$ and $0$, supersymmetric
extreme black holes arising from string compactifications were found.
Moreover, from both the spacetime solutions \bjoachim\ and
supersymmetries \bsusy\ point of view, these four solutions
can be interpreted as bound states of $1,2,3$ and $4$ distinct
$a=\sqrt{3}$
black holes, respectively,
the latter corresponding to maximally supersymmetric
($N=4$ supersymmetry in an $N=8$, $D=4$ theory)
Kaluza-Klein \kal\ or $H$-monopole \hmono\ solutions with flat
metric on moduli space \rust.
For example, the $a=0$ Reissner-Nordstr\"om
black hole arises as a bound state of two $T$-dual
pairs of electric/magnetic $a=\sqrt{3}$ black holes, each pair
producing
an $a=1$ black hole. Again, this feature extend quite naturally to
arbitrary supersymmetric $p$-branes in any dimension, as well as to
non-extremal, non-supersymmetric $p$-branes \imirjam.

This compositeness feature lies at the heart of the recent success
of string theory in reproducing the Beckenstein-Hawking formula
for the entropy of black holes \refs{\exfive\exfour{--}\nonex}.
In particular, for a given $p$-brane arising as a bound state
or as intersections of basic constituent $p$-branes of charge
$Q_1$,$Q_2$,$\ldots$,$Q_n$, the area law for the entropy yields
\eqn\entropy{S=A/4G = 2\pi \sqrt{Q_1 Q_2\ldots Q_n},}
which agrees, for large $Q_i$, with
the microscopic entropy formula obtained by counting
string states. This was first seen for five-dimensional extremal
black holes in \exfive\ and subsequently for four-dimensional
extremal black holes in \exfour. Analogous results for near-extremal
black holes were obtained in \nonex, which seems to indicate that
this sort of factorization is not a property of supersymmetry alone,
although it is only for supersymmetric solutions that one can
invoke non-renormalization theorems to protect the counting
of states in going from the perturbative state-counting picture
to the non-perturbative black hole picture.

The question then arises as to whether a similar property holds
for arbitrary non-extremal black holes. Such a formula was, in fact,
found from the $p$-brane picture in \nonex, where, however, it was
noted that the corresponding $D$-brane counting argument was unknown.
More recently, it was argued in \fat\
that the counting arguments relating perturbative states to black
holes break down for ``fat'' black holes, {\it i.e.} large black
holes
far from extremality (such as the Schwarzschild black hole). However,
this does not in itself rule out the possibility of an analogous,
if not identical, compositeness
feature in the general case which will once again recover the
Beckenstein-Hawking formula and confirm an important success
of string theory as a theory of quantum gravity.

\noindent

\vfil\eject
\listrefs
\bye